**4**

# On the Modelling of Spur and Helical Gear Dynamic Behaviour

Philippe Velex
*University of Lyon, INSA Lyon, LaMCoS UMR CNRS,*
*France*

## 1. Introduction

This chapter is aimed at introducing the fundamentals of spur and helical gear dynamics. Using three-dimensional lumped models and a thin-slice approach for mesh elasticity, the general equations of motion for single-stage spur or helical gears are presented. Some particular cases including the classic one degree-of-freedom model are examined in order to introduce and illustrate the basic phenomena. The interest of the concept of transmission errors is analysed and a number of practical considerations are deduced. Emphasis is deliberately placed on analytical results which, although approximate, allow a clearer understanding of gear dynamics than that provided by extensive numerical simulations. Some extensions towards continuous models are presented.

## 2. Nomenclature

$b$ : face width

$C_m, C_r$ : pinion, gear torque

$e(M)$, $E_{MAX}(t)$ : composite normal deviation at $M$, maximum of $e(M)$ at time $t$.

$E, E^*$ : actual and normalized depth of modification at tooth tips

$\mathbf{F_e}(t) = \int_{L(t,\mathbf{q})} k(M)\delta e(M)\mathbf{V}(\mathbf{M})dM$ : time-varying, possibly non-linear forcing term associated with tooth shape modifications and errors

$\mathbf{G} = \mathbf{V}_0 \mathbf{V}_0^T$

$H(x)$ : unit Heaviside step function ( $H(x)=1$ *if* $x>1; H(x)=0$ *otherwise* )

$k_m$, $k(t,\mathbf{q})$ : average and time-varying, non-linear mesh stiffness

$k(t) = k_m(1+\alpha\varphi(t))$, linear time-varying mesh stiffness

$k_0$ : mesh stiffness per unit of contact length

$k(M)$, mesh stiffness per unit of contact length at $M$

$k_{\Phi_p}$ : modal stiffness associated with ( $\omega_p$, $\mathbf{\Phi_p}$ )

$[\mathbf{K_G}(t)] = \int_{L(t,\mathbf{q})} k(M)\mathbf{V}(\mathbf{M})\ \mathbf{V}(\mathbf{M})^T dM$ : time-varying, possibly non-linear gear mesh stiffness matrix

$L(t, \mathbf{q})$ : time-varying, possibly non-linear, contact length

$L_m = \varepsilon_\alpha \dfrac{b}{\cos \beta_b}$ : average contact length

$\hat{m} = \dfrac{I_{02} I_{01}}{Rb_1^2 I_{02} + Rb_2^2 I_{01}}$ : equivalent mass

$m_{\Phi p}$ : modal mass associated with ($\omega_p$, $\mathbf{\Phi_p}$)

$\mathbf{n_1}$ : outward unit normal vector with respect to pinion flanks

$NLTE$ : no-load transmission error

$O_1, O_2$ : pinion, gear centre

$Pb_a$ : apparent base pitch

$Rb_1, Rb_2$ : base radius of pinion, of gear

$(\mathbf{s}, \mathbf{t}, \mathbf{z})$ : coordinate system attached to the pinion-gear centre line, see Figs. 1&2

$T_m$ : mesh period.

$TE$, $TE_S$ : transmission error, quasi-static transmission error under load

$\mathbf{V}(M), \mathbf{V_0}$, structural vector, averaged structural vector

$\mathbf{W}$ : projection vector for the expression of transmission error, see (44-1)

$(\mathbf{X}, \mathbf{Y}, \mathbf{z})$ : coordinate system associated with the base plane, see Fig. 2

$\mathbf{X_0} = \bar{\mathbf{K}}^{-1} \mathbf{F_0}$ : static solution with averaged mesh stiffness (constant)

$\mathbf{X_S}$, $\mathbf{X_D}$ $\mathbf{X}$ : quasi-static, dynamic and total (elastic) displacement vector (time-dependent)

$Z_1, Z_2$ : tooth number on pinion, on gear

$\alpha$ : small parameter representative of mesh stiffness variations, see (30)

$\alpha_p$ : apparent pressure angle

$\beta_b$ : base helix angle

$\delta_m = \dfrac{F_S}{k_m} = \mathbf{V^T X_0}$ : static mesh deflection with average mesh stiffness

$\delta e(M) = E_{MAX}(t) - e(M)$ : instantaneous initial equivalent normal gap at $M$

$\Delta(M)$ : mesh deflection at point $M$

$\varepsilon_\alpha$ : theoretical profile contact ratio

$\varepsilon_\beta$ : overlap contact ratio

$\Lambda = \dfrac{Cm}{Rb_1 b k_0}$ , deflection of reference

$\mathbf{\Phi_p}$ : $p^{th}$ eigenvector of the system with constant averaged stiffness matrix

$\varsigma_P$ : damping factor associated with the $p^{th}$ eigenfrequency

$\Gamma$ : dimensionless extent of profile modification (measured on base plane)

$\tau = \dfrac{t}{T_m}$ , dimensionless time

$\omega_p$ : $p^{th}$ eigenfrequency of the system with constant averaged stiffness matrix

$\varpi_{pn} = \dfrac{\omega_p}{n\Omega_1}$ , dimensionless eigenfrequency

$\Omega_1, \Omega_2$ : pinion, gear angular velocity

$\overline{\mathbf{A}}$ : vector $\mathbf{A}$ completed by zeros to the total system dimension

$(\bullet)^* = \dfrac{(\bullet)}{\delta_m}$ , normalized displacement with respect to the average static mesh deflection

$(\hat{\bullet}) = \dfrac{(\bullet)}{k_m}$ , normalized stiffness with respect to the average mesh stiffness

## 3. Three-dimensional lumped parameter models of spur and helical gears

### 3.1 Rigid-body rotations – State of reference

It is well-known that the speed ratio for a pinion-gear pair with perfect involute spur or helical teeth is constant as long as deflections can be neglected. However, shape errors are present to some extent in all gears as a result of machining inaccuracy, thermal distortions after heat treatment, etc. Having said this, some shape modifications from ideal tooth flanks are often necessary (profile and/or lead modifications, topping) in order to compensate for elastic or thermal distortions, deflections, misalignments, positioning errors, etc. From a simulation point of view, rigid-body rotations will be considered as the references in the vicinity of which, small elastic displacements can be superimposed. It is therefore crucial to characterise rigid-boy motion transfer between a pinion and a gear with tooth errors and/or shape modifications. In what follows, *e(M)* represents the equivalent normal deviation at the potential point of contact M (sum of the deviations on the pinion and on the gear) and is conventionally positive for an excess of material and negative when, on the contrary, some material is removed from the ideal geometry. For rigid-body conditions (or alternatively under no-load), contacts will consequently occur at the locations on the contact lines where *e(M)* is maximum and the velocity transfer from the pinion to the gear is modified compared with ideal gears such that:

$$(Rb_1 \Omega_1 + Rb_2 \Omega_2)\cos\beta_b + \dfrac{dE_{MAX}(t)}{dt} = 0 \qquad (1)$$

where $E_{MAX}(t) = \max_M(e(M))$ with $\max_M()$, maximum over all the potential point of contact at time *t*

The difference with respect to ideal motion transfer is often related to the notion of no-load transmission error *NLTE* via:

$$\dfrac{d}{dt}(NLTE) = Rb_1 \Omega_1 + Rb_2 \Omega_2 = -\dfrac{1}{\cos\beta_b}\dfrac{dE_{MAX}(t)}{dt} \qquad (2)$$

Using the Kinetic Energy Theorem, the rigid-body dynamic behaviour for frictionless gears is controlled by:

$$J_1 \Omega_1 \dot{\Omega}_1 + J_2 \Omega_2 \dot{\Omega}_2 = C_m \Omega_1 + C_r \Omega_2 \qquad (3)$$

with $J_1, J_2$: the polar moments of inertia of the pinion shaft line and the gear shaft line respectively. $C_m, C_r$: pinion and gear torques.

The system with 4 unknowns ($\Omega_1, \Omega_2, C_m, C_r$) is characterised by equations (2) - (3) only, and 2 parameters have to be imposed.

### 3.2 Deformed state – Principles

Modular models based on the definition of gear elements (pinion-gear pairs), shaft elements and lumped parameter elements (mass, inertia, stiffness) have proved to be effective in the simulation of complex gear units (Küçükay, 1987), (Baud & Velex, 2002). In this section, the theoretical foundations upon which classic gear elements are based are presented and the corresponding elemental stiffness and mass matrices along with the possible elemental forcing term vectors are derived and explicitly given. The simplest and most frequently used 3D representation corresponds to the pinion-gear model shown in Figure 1. Assuming that the geometry is not affected by deflections (small displacements hypothesis) and provided that mesh elasticity (and to a certain extent, gear body elasticity) can be transferred onto the base plane, a rigid-body approach can be employed. The pinion and the gear can therefore be assimilated to two rigid cylinders with 6 degrees of freedom each, which are connected by a stiffness element or a distribution of stiffness elements (the discussion of the issues associated with damping and energy dissipation will be dealt with in section 4.3). From a physical point of view, the 12 degrees of freedom of a pair represent the generalised displacements of i) traction: $u_1, u_2$ (axial displacements), ii) bending: $v_1, w_1, v_2, w_2$ (translations in two perpendicular directions of the pinion/gear centre), $\varphi_1, \psi_1, \varphi_2, \psi_2$ (bending rotations which can be assimilated to misalignment angles) and finally, iii) torsion:

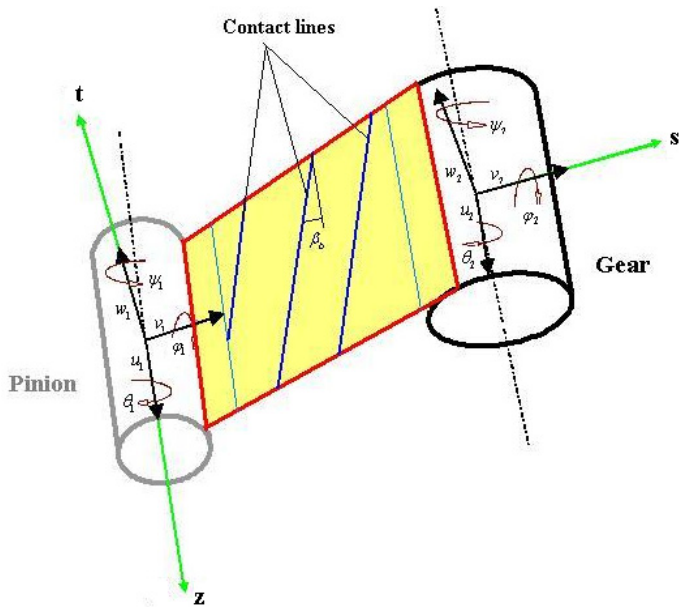

Fig. 1. A 3D lumped parameter model of pinion-gear pair.

$\theta_1, \theta_2$ which are small angles associated with deflections superimposed on rigid-body rotations $\Theta_1 = \int_0^t \Omega_1(\sigma) d\sigma$ (pinion) and $\Theta_2 = \int_0^t \Omega_2(\sigma) d\sigma$ (gear). Following Velex and Maatar (1996), screws of infinitesimal displacements are introduced whose co-ordinates for solid $k$ (*conventionally k=1 for the pinion, k=2 for the gear*) can be expressed in two privileged coordinate systems: i) $(\mathbf{s}, \mathbf{t}, \mathbf{z})$ such that $\mathbf{z}$ is in the shaft axis direction (from the motor to the load machine), $\mathbf{s}$ is in the centre-line direction from the pinion centre to the gear centre and $\mathbf{t} = \mathbf{z} \times \mathbf{s}$ (Fig. 1) or, ii) $(\mathbf{X}, \mathbf{Y}, \mathbf{z})$ attached to the base plane (Fig. 1):

$$\{S_k\} \begin{Bmatrix} \mathbf{u_k}(\mathbf{O_k}) = v_k \mathbf{s} + w_k \mathbf{t} + u_k \mathbf{z} \\ \mathbf{\omega_k} = \varphi_k \mathbf{s} + \psi_k \mathbf{t} + \theta_k \mathbf{z} \end{Bmatrix} \text{ or } \begin{Bmatrix} \mathbf{u_k}(\mathbf{O_k}) = V_k \mathbf{X} + W_k \mathbf{Y} + u_k \mathbf{z} \\ \mathbf{\omega_k} = \Phi_k \mathbf{X} + \Psi_k \mathbf{Y} + \theta_k \mathbf{z} \end{Bmatrix} \quad k=1,2 \qquad (4)$$

where $O_1, O_2$ are the pinion and gear centres respectively

### 3.3 Deflection at a point of contact – Structural vectors for external gears

Depending on the direction of rotation, the direction of the base plane changes as illustrated in Figure 2 where the thicker line corresponds to a positive rotation of the pinion and the finer line to a negative pinion rotation about axis $(O_1, \mathbf{z})$.

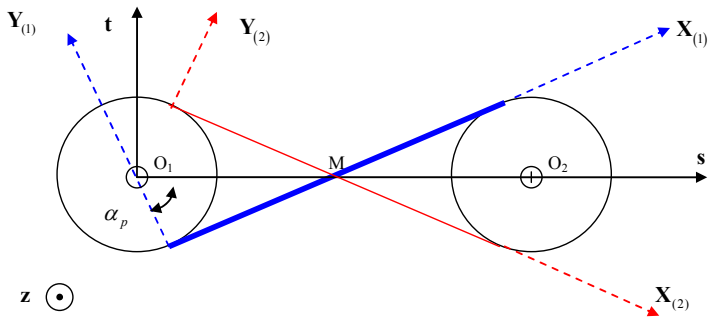

Fig. 2. Directions of rotation and planes (lines) of action. (*the thicker line corresponds to a positive rotation of pinion*)

For a given helical gear, the sign of the helix angle on the base plane depends also on the direction of rotation and, here again; two configurations are possible as shown in Figure 3.

Since a rigid-body mechanics approach is considered, contact deflections correspond to the interpenetrations of the parts which are deduced from the contributions of the degrees-of-freedom and the initial separations both measured in the normal direction with respect to the tooth flanks. Assuming that all the contacts occur in the theoretical base plane (or plane of action), the normal deflection $\Delta(M)$ at any point $M$, potential point of contact, is therefore expressed as:

$$\Delta(M) = \mathbf{u_1}(\mathbf{M}) \cdot \mathbf{n_1} - \mathbf{u_2}(\mathbf{M}) \cdot \mathbf{n_1} - \delta e(M) \qquad (5)$$

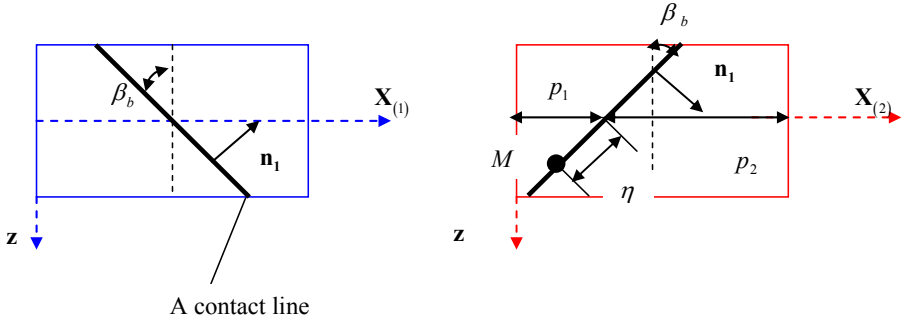

Fig. 3. Helix angles on the base plane.

where $\delta e(M) = \max_M(e(M)) - e(M)$ is the equivalent initial normal gap at $M$ caused by tooth modifications and/or errors for example, $\mathbf{n_1}$ is the outward unit normal vector to pinion tooth flanks (Fig.3)

Using the shifting property of screws, one obtains the expression of $\Delta(M)$ in terms of the screw co-ordinates as:

$$\Delta(M) = \mathbf{u_1}(\mathbf{O_1}).\mathbf{n_1} + (\boldsymbol{\omega_1} \times \mathbf{O_1M}).\mathbf{n_1} - \mathbf{u_2}(\mathbf{O_2}).\mathbf{n_1} - (\boldsymbol{\omega_2} \times \mathbf{O_2M}).\mathbf{n_1} - \delta e(M) \quad (6)$$

which is finally expressed as:

$$\Delta(M) = \begin{bmatrix} \mathbf{n_1} \\ \mathbf{O_1M} \times \mathbf{n_1} \\ -\mathbf{n_1} \\ -\mathbf{O_2M} \times \mathbf{n_1} \end{bmatrix}^T \begin{bmatrix} \mathbf{u_1}(\mathbf{O_1}) \\ \boldsymbol{\omega_1} \\ \mathbf{u_2}(\mathbf{O_2}) \\ \boldsymbol{\omega_2} \end{bmatrix} - \delta e(M) \quad (7)$$

or, in a matrix form:

$$\Delta(M) = \mathbf{V(M)}^T \mathbf{q} - \delta e(M) \quad (8)$$

where $\mathbf{V(M)}$ is a structural vector which accounts for gear geometry (Küçükay, 1987) and $\mathbf{q}$ is the vector of the pinion-gear pair degrees of freedom (*superscript T refers to the transpose of vectors and matrices*)

The simplest expression is that derived in the $(\mathbf{X}, \mathbf{Y}, \mathbf{z})$ coordinate system associated with the base plane leading to:

$$\mathbf{V(M)}^T = \langle \cos\beta_b, \ 0, \ \varepsilon\sin\beta_b, \ -\zeta\varepsilon Rb_1\sin\beta_b, \ \eta - \varepsilon p_1\sin\beta_b, \ \zeta Rb_1\cos\beta_b,$$
$$-\cos\beta_b, \ 0, \ -\varepsilon\sin\beta_b, \ -\zeta\varepsilon Rb_2\sin\beta_b, \ -[\eta + \varepsilon p_2\sin\beta_b], \ \zeta Rb_2\cos\beta_b \ \rangle$$

$$\mathbf{q}^T = \langle V_1 \ W_1 \ u_1 \ \Phi_1 \ \Psi_1 \ \theta_1 \ V_2 \ W_2 \ u_2 \ \Phi_2 \ \Psi_2 \ \theta_2 \rangle \quad (9)$$

where $Rb_1, Rb_2$ are the pinion, gear base radii; $\beta_b$ is the base helix angle (*always considered as positive in this context*); $p_1, p_2, \eta$ are defined in Figure 3; $\varepsilon = \pm 1$ depending on the sign of the helix angle; $\zeta = +1$ for a positive rotation of the pinion and $\zeta = -1$ for a negative rotation of the pinion.

An alternative form of interest is obtained when projecting in the $(\mathbf{s}, \mathbf{t}, \mathbf{z})$ frame attached to the pinion-gear centre line:

$$\mathbf{V}(M)^T = \langle \cos\beta_b \sin\alpha_p, \quad \zeta\cos\beta_b\cos\alpha_p, \quad \varepsilon\sin\beta_b, \quad -\zeta\varepsilon Rb_1 \sin\beta_b \sin\alpha_p - \zeta(\eta - \varepsilon p_1 \sin\beta_b)\cos\alpha_p$$
$$\left[-\varepsilon Rb_1 \sin\beta_b \cos\alpha_p + (\eta - \varepsilon p_1 \sin\beta_b)\sin\alpha_p\right], \quad \zeta Rb_1 \cos\beta_b, \quad -\cos\beta_b \sin\alpha_p, \quad -\zeta\cos\beta_b \cos\alpha_p,$$
$$-\varepsilon\sin\beta_b, \quad -\zeta\varepsilon Rb_2 \sin\beta_b \sin\alpha_p + \zeta(\eta + \varepsilon p_2 \sin\beta_b)\cos\alpha_p$$
$$-\left[\varepsilon Rb_2 \sin\beta_b \cos\alpha_p + (\eta + \varepsilon p_2 \sin\beta_b)\sin\alpha_p\right], \quad \zeta Rb_2 \cos\beta_b \rangle$$

$$\mathbf{q}^T = \langle v_1 \quad w_1 \quad u_1 \quad \varphi_1 \quad \psi_1 \quad \theta_1 \quad v_2 \quad w_2 \quad u_2 \quad \varphi_2 \quad \psi_2 \quad \theta_2 \rangle \tag{10}$$

### 3.4 Mesh stiffness matrix and forcing terms for external gears

For a given direction of rotation, the usual contact conditions in gears correspond to single-sided contacts between the mating flanks which do not account for momentary tooth separations which may appear if dynamic displacements are large (of the same order of magnitude as static displacements). A review of the mesh stiffness models is beyond the scope of this chapter but one usually separates the simulations accounting for elastic convection (i.e., the deflection at one point $M$ depends on the entire load distribution on the tooth or all the mating teeth (Seager, 1967)) from the simpler (and classic) thin-slice approach (the deflection at point $M$ depends on the load at the same point only). A discussion of the limits of this theory can be found in Haddad (1991), Ajmi & Velex (2005) but it seems that, for solid gears, it is sufficiently accurate as far as dynamic phenomena such as critical speeds are considered as opposed to exact load or stress distributions in the teeth which are more dependent on local conditions. Neglecting contact damping and friction forces compared with the normal elastic components on tooth flanks, the elemental force transmitted from the pinion onto the gear at one point of contact $M$ reads:

$$d\mathbf{F}_{1/2}(M) = k(M)\Delta(M)dM\,\mathbf{n}_1 \tag{11}$$

with $k(M)$: mesh stiffness at point $M$ per unit of contact length

The resulting total mesh force and moment at the gear centre $O_2$ are deduced by integrating over the time-varying and possibly deflection-dependent contact length $L(t,\mathbf{q})$ as:

$$\{F_{1/2}\}\begin{cases}\mathbf{F}_{1/2} = \int\limits_{L(t,\mathbf{q})} k(M)\Delta(M)dM\,\mathbf{n}_1 \\ \mathbf{M}_{1/2}(\mathbf{O_2}) = \int\limits_{L(t,\mathbf{q})} k(M)\Delta(M)\mathbf{O_2M}\times\mathbf{n}_1\,dM\end{cases} \tag{12-1}$$

Conversely the mesh force wrench at the pinion centre $O_1$ is:

$$\{F_{2/1}\}\begin{cases} \mathbf{F_{2/1}} = -\int\limits_{L(t,\mathbf{q})} k(M)\Delta(M)dM\,\mathbf{n_1} \\ \mathbf{M_{2/1}}(\mathbf{O_1}) = -\int\limits_{L(t,\mathbf{q})} k(M)\Delta(M)\mathbf{O_1M}\times\mathbf{n_1}\,dM \end{cases} \quad (12\text{-}2)$$

The mesh inter-force wrench can be deduced in a compact form as:

$$\{F_M\}\begin{cases}\{F_{2/1}\}\\ \{F_{1/2}\}\end{cases} = -\int\limits_{L(t,\mathbf{q})} k(M)\Delta(M)\mathbf{V}(\mathbf{M})\,dM \quad (13)$$

and introducing the contact normal deflection $\Delta(M) = \mathbf{V}(\mathbf{M})^T\mathbf{q} - \delta e(M)$ finally leads to:

$$\{F_M\} = -[\mathbf{K_G}(t)]\mathbf{q} + \mathbf{F_e}(t) \quad (14)$$

where $[\mathbf{K_G}(t)] = \int\limits_{L(t,\mathbf{q})} k(M)\mathbf{V}(\mathbf{M})\,\mathbf{V}(\mathbf{M})^T\,dM$ is the time-varying gear mesh stiffness matrix

$\mathbf{F_e}(t) = \int\limits_{L(t,\mathbf{q})} k(M)\delta e(M)\mathbf{V}(\mathbf{M})\,dM$ is the excitation vector associated with tooth shape modifications and errors

### 3.5 Mass matrix of external gear elements – Additional forcing (inertial) terms

For solid $k$ (pinion or gear), the dynamic sum with respect to the inertial frame can be expressed as:

$$\mathbf{\Sigma_k^0} = m_k\left[\left(\ddot{v}_k - e_k\dot{\Omega}_k\sin\Theta_k - e_k\Omega_k^2\cos\Theta_k\right)\mathbf{s} + \left(\ddot{w}_k + e_k\dot{\Omega}_k\cos\Theta_k - e_k\Omega_k^2\sin\Theta_k\right)\mathbf{t} + \ddot{u}_k\mathbf{z}\right] \quad (15)$$

where $m_k$ and $e_k$ are respectively the mass and the eccentricity of solid $k$

A simple expression of the dynamic moment at point $O_k$ can be obtained by assuming that $O_k$ is the centre of inertia of solid $k$ and neglecting gyroscopic components (complementary information can be found in specialised textbooks on rotor dynamics (see for instance (Lalanne & Ferraris, 1998)):

$$\boldsymbol{\delta}_\mathbf{k}^0(O_k) \cong I_k\ddot{\phi}_k\mathbf{s} + I_k\ddot{\psi}_k\mathbf{t} + I_{0k}\left(\dot{\Omega}_k + \ddot{\theta}_k\right)\mathbf{z} \quad (16)$$

where $I_k$ is the cross section moment of inertia and $I_{0k}$ is the polar moment of solid k

Using the same DOF arrangement as for the stiffness matrices, a mass matrix for the pinion-gear system can be deduced as (note that the same mass matrix is obtained in the $(\mathbf{X},\mathbf{Y},\mathbf{z})$ coordinate system):

$$[\mathbf{M_G}] = \mathbf{diag}(m_1,m_1,m_1,I_1,I_1,I_{01},m_2,m_2,m_2,I_2,I_2,I_{02}) \quad (17\text{-}1)$$

along with a forcing term associated with inertial forces (whose expression in $(\mathbf{X},\mathbf{Y},\mathbf{z})$ has the same form on the condition that angles $\Theta_{1,2}$ are measured from $\mathbf{X}$ and $\mathbf{Y}$):

$$\mathbf{F_G}(t) = \big\langle m_1 e_1 \big(\dot{\Omega}_1 \sin\Theta_1 + \Omega_1^2 \cos\Theta_1\big) \quad -m_1 e_1 \big(\dot{\Omega}_1 \cos\Theta_1 - \Omega_1^2 \sin\Theta_1\big) \quad 0 \quad 0 \quad 0 \quad -I_{01}\dot{\Omega}_1 \\ m_2 e_2 \big(\dot{\Omega}_2 \sin\Theta_2 + \Omega_2^2 \cos\Theta_2\big) \quad -m_2 e_2 \big(\dot{\Omega}_2 \cos\Theta_2 - \Omega_2^2 \sin\Theta_2\big) \quad 0 \quad 0 \quad 0 \quad -I_{02}\dot{\Omega}_2 \big\rangle \tag{17-2}$$

### 3.6 Usual simplifications

Examining the components of the structural vectors in (9) and (10), it can be noticed that most of them are independent of the position of the point of contact M with the exception of those related to bending slopes $\Psi_{1,2}$ or $\varphi_{1,2}, \psi_{1,2}$. Their influence is usually discarded especially for narrow-faced gears so that the mesh stiffness matrix can be simplified as:

$$\big[\mathbf{K_G}(t)\big] \cong \int_{L(t,\mathbf{q})} k(M)dM \; \mathbf{V}_0 \mathbf{V}_0^T = k(t,\mathbf{q})\mathbf{G} \tag{18}$$

where $\mathbf{V}_0$ represents an average structural vector and $k(t,\mathbf{q})$ is the time-varying, possibly non-linear, mesh stiffness function (scalar) which plays a fundamental role in gear dynamics.

### 3.6.1 Classic one-DOF torsional model

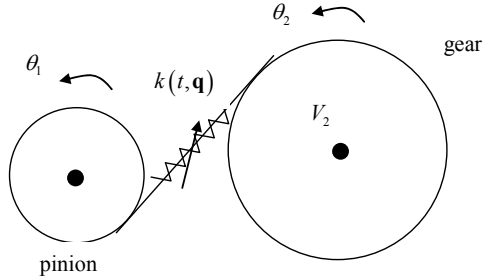

Fig. 4. Basic torsional model.

Considering the torsional degrees-of-freedom only (Figure 4), the structural vector reads (keeping solely the non-zero components):

$$\mathbf{V}(\mathbf{M}) = \mathbf{V_0} = \begin{bmatrix} \zeta\,Rb_1 \\ \zeta\,Rb_2 \end{bmatrix}\cos\beta_b \tag{19}$$

and the following differential system is derived $\big(\zeta^2 = 1\big)$:

$$\begin{bmatrix} I_{01} & 0 \\ 0 & I_{02} \end{bmatrix}\begin{bmatrix} \ddot{\theta}_1 \\ \ddot{\theta}_2 \end{bmatrix} + k(t,\theta_1,\theta_2)\cos^2\beta_b \begin{bmatrix} Rb_1^2 & Rb_1 Rb_2 \\ Rb_1 Rb_2 & Rb_2^2 \end{bmatrix}\begin{bmatrix} \theta_1 \\ \theta_2 \end{bmatrix} = \\ = \begin{bmatrix} Cm \\ Cr \end{bmatrix} + \int_{L(t,\mathbf{q})} k(M)\delta e(M)dM \begin{bmatrix} \zeta\,Rb_1 \\ \zeta\,Rb_2 \end{bmatrix}\cos\beta_b - \begin{bmatrix} I_{01}\dot{\Omega}_1 \\ I_{02}\dot{\Omega}_2 \end{bmatrix} \tag{20}$$

Note that the determinant of the stiffness matrix is zero which indicates a rigid-body mode (the mass matrix being diagonal). After multiplying the first line in (20) by $Rb_1 I_{02}$, the second line by $Rb_2 I_{01}$, adding the two equations and dividing all the terms by $\left(I_{10} Rb_2^2 + I_{20} Rb_1^2\right)$, the semi-definite system (20) is transformed into the differential equation:

$$\widehat{m}\ddot{x} + k(t,x)x = F_t + \zeta \cos\beta_b \int_{L(t,x)} k(M)\delta e(M)dM - \kappa \frac{d^2}{dt^2}(NLTE) \tag{21}$$

With $x = Rb_1\theta_1 + Rb_2\theta_2$, relative apparent displacement

$\widehat{m} = \dfrac{I_{02} I_{01}}{Rb_1^2 I_{02} + Rb_2^2 I_{01}}$, equivalent mass

$\kappa = \Omega_1^2 \dfrac{I_{02}}{Rb_2^2}$ when the pinion speed $\Omega_1$ and the output torque $C_r$ are supposed to be constant.

### 3.6.2 A simple torsional-flexural model for spur gears

The simplest model which accounts for torsion and bending in spur gears is shown in Figure 5. It comprises 4 degrees of freedom, namely: 2 translations in the direction of the line of action $V_1, V_2$ (at pinion and gear centres respectively) and 2 rotations about the pinion and gear axes of rotation $\theta_1, \theta_2$. Because of the introduction of bending DOFs, some supports (bearing/shaft equivalent stiffness elements for instance) must be added.

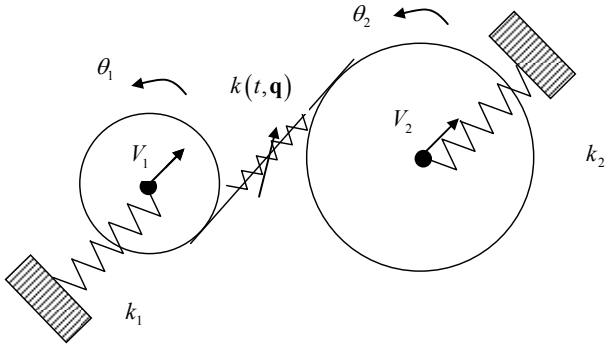

Fig. 5. Simplified torsional-flexural spur gear model.

The general expression of the structural vector $\mathbf{V}(M)$ (9) reduces to:

$$\mathbf{V_0}^T = \langle 1 \quad \zeta Rb_1 \quad -1 \quad \zeta Rb_2 \rangle \tag{22}$$

Re-writing the degree of freedom vector as $\mathbf{q}^{*T} = \langle v_1 \quad Rb_1\theta_1 \quad v_2 \quad Rb_2\theta_2 \rangle$, the following parametrically excited differential system is obtained for linear free vibrations:

$$\mathbf{M}\ddot{\mathbf{q}}^* + \mathbf{K}(t)\mathbf{q}^* = \mathbf{0} \tag{23-1}$$

$$\mathbf{M} = \begin{bmatrix} m_1 & & & \\ & I_{01}/Rb_1^2 & & \\ & & m_2 & \\ & & & I_{02}/Rb_2^2 \end{bmatrix}; \quad \mathbf{K}(t) = \begin{bmatrix} k(t)+k_1 & \zeta k(t) & -k(t) & \zeta k(t) \\ & k(t) & -\zeta k(t) & k(t) \\ & & k(t)+k_2 & -\zeta k(t) \\ & & & k(t) \end{bmatrix} \quad (23\text{-}2)$$

*Remark:* The system is ill-conditioned since rigid-body rotations are still possible (no unique static solution). In the context of 3D models with many degrees of freedom, it is not interesting to solve for the normal approach $Rb_1\theta_1 + Rb_2\theta_1$ as is done for single DOF models. The problem can be resolved by introducing additional torsional stiffness element(s) which can represent shafts; couplings etc. thus eliminating rigid-body rotations.

## 4. Mesh stiffness models – Parametric excitations

### 4.1 Classic thin-slice approaches

From the results in section 2-5, it can be observed that, in the context of gear dynamic simulations, the mesh stiffness function defined as $k(t,\mathbf{q}) = \int_{L(t,\mathbf{q})} k(M)dM$ plays a key role. This function stems from a 'thin-slice' approach whereby the contact lines between the mating teeth are divided in a number of independent stiffness elements (with the limiting case presented here of an infinite set of non-linear time-varying elemental stiffness elements) as schematically represented in Figure 6.

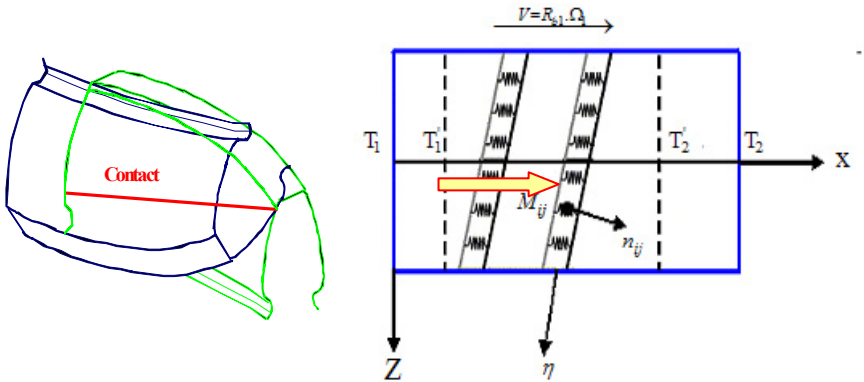

Fig. 6. 'Thin-slice' model for time-varying mesh stiffness.

Since the positions of the teeth (and consequently the contact lines) evolve with time (or angular positions), the profiles slide with respect to each other and the stiffness varies because of the contact length and the individual tooth stiffness evolutions. The definition of mesh stiffness has generated considerable interest but mostly with the objective of calculating accurate static tooth load distributions and stress distributions. It has been shown by Ajmi and Velex (2005) that a classic 'thin-slice' model is sufficient for dynamic calculations as long as local disturbances (especially near the tooth edges) can be ignored. In this context, Weber and Banascheck (1953) proposed a analytical method of calculating tooth deflections of spur gears by superimposing displacements which arise from i) the contact between the teeth, ii) the tooth itself considered as a beam and, iii) the gear body (or foundation) influence. An analytical expression of the contact compliance was obtained using the 2D Hertzian theory for cylinders in contact which is singular as far as the normal approach between the parts (contact deflection) is concerned. The other widely-used formulae for tooth contact deflection comprise the analytical formula of Lundberg (1939), the approximate Hertzian approach originally used at Hamilton Standard (Cornell, 1981) and the semi-empirical formula developed by Palmgren (1959) for rollers. The tooth bending radial and tangential displacements were derived by equating the work produced by one individual force acting on the tooth profile and the strain energy of the tooth assimilated to a cantilever of variable thickness. Extensions and variants of the methodology were introduced by Attia (1964), Cornell (1981) and O'Donnell (1960, 1963) with regard to the foundation effects. Gear body contributions were initially evaluated by approximating them as part of an elastic semi-infinite plane loaded by the reactions at the junction with the tooth. A more accurate expression for this base deflection has been proposed by Sainsot *et al.* (2004) where the gear body is simulated by an elastic annulus instead of a half-plane. Figure 7 shows two examples of mesh stiffness functions (no contact loss) calculated by combining Weber's and Lundberg's results for a spur and a helical gear example. It can be observed that the stiffness fluctuations are stronger in the case of conventional spur gears compared with helical gears for which the contact variations between the teeth are smoother.

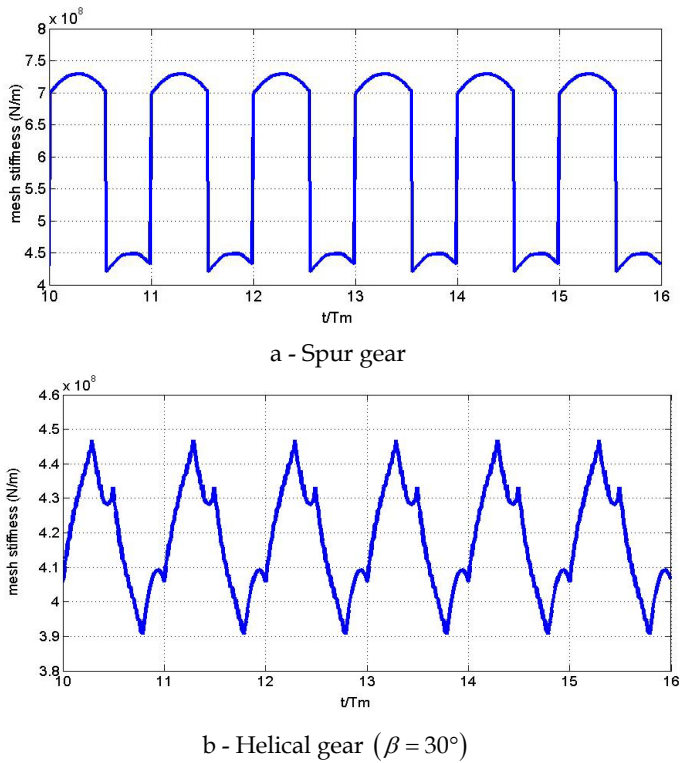

a - Spur gear

b - Helical gear $(\beta = 30°)$

Fig. 7. Examples of mesh stiffness functions for errorless gears.

Although the results above are based on simplified bi-dimensional approaches, they are still widely used in gear design. For example, the mesh stiffness formulae in the ISO standard 6336 stem from Weber's analytical formulae which were modified to bring the values in closer agreement with the experimental results. Another important simplification brought by the ISO formulae is that the mesh stiffness per unit of contact length $k_0$ is considered as approximately constant so that the following approximation can be introduced:

$$\int_{L(t,\mathbf{q})} k(M)dM \cong k_0 \int_{L(t,\mathbf{q})} dM = k_0 L(t,\mathbf{q}) \qquad (24)$$

where $L(t,\mathbf{q})$ is the time-varying (possibly non-linear) contact length.

## 4.2 Contact length variations for external spur and helical gears

Considering involute profiles, the contact lines in the base plane are inclined by the base helix angle $\beta_b$ (Figure 8) which is nil for spur gears. All contact lines are spaced by integer multiples of the apparent base pitch $Pb_a$ and, when the pinion and the gear rotate, they all undergo a translation in the $\mathbf{X}$ direction at a speed equal to $Rb_1 \Omega_1$.

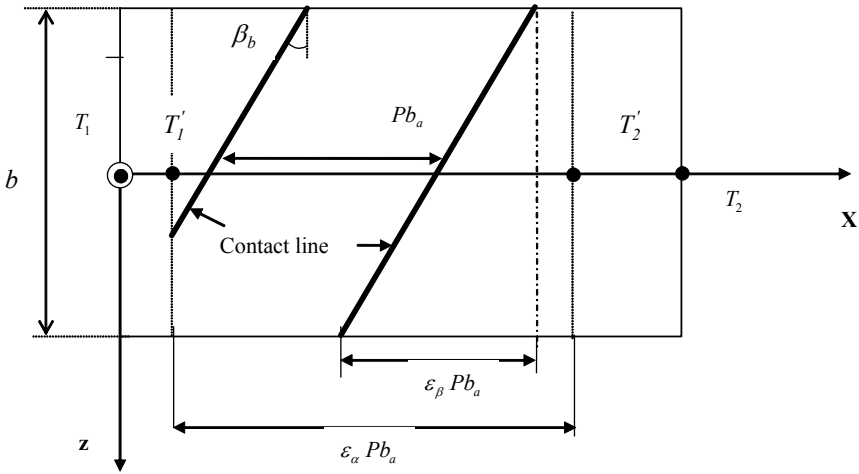

Fig. 8. Base plane and contact lines ($b$: face width; $\mathbf{Z}$: axial direction (direction of the axes of rotation); $T_1, T_2$: points of tangency on pinion and gear base circles and $T_1^{'}, T_2^{'}$: limits of the contact area on base plane).

It transpires from this geometrical representation that the total length of contact between the pinion and the gear is likely to vary with time and, based on the simple stiffness equation (24), that mesh stiffness is time-varying and, consequently, contributes to the system excitation via parametric excitations.

The extent of action on the base plane is an important property measured by the contact ratio $\varepsilon_\alpha$ which, in simple terms, represents the 'average number' of tooth pairs in contact (possibly non integer) and is defined by:

$$\varepsilon_\alpha = \frac{T_1'T_2'}{Pb_a} = \frac{\sqrt{Ra_1^2 - Rb_1^2} + \sqrt{Ra_2^2 - Rb_2^2} - E\sin\alpha_p}{\pi m \cos\alpha_p} \quad (25\text{-}1)$$

with $Ra_1, Ra_2$: external radius of pinion, of gear; $Rb_1, Rb_2$: base radius of pinion, of gear; $E = \|\vec{O_1O_2}\|$: centre distance

In the case of helical gears, the overlap due to the helix is taken into account by introducing the overlap ratio $\varepsilon_\beta$ defined as:

$$\varepsilon_\beta = \frac{b\tan\beta_b}{Pb_a} = \frac{1}{\pi}\frac{b}{m}\frac{\tan\beta_b}{\cos\alpha_p} \quad (25\text{-}2)$$

and the sum $\varepsilon = \varepsilon_\alpha + \varepsilon_\beta$ is defined as the total contact ratio.

Introducing the dimensionless time $\tau = \dfrac{t}{T_m}$ where $T_m = \dfrac{Pb_a}{Rb_1\Omega_1}$ is the mesh period i.e. the time needed for a contact line to move by a base pitch on the base plane, a closed form expression of the contact length $L(\tau)$ for ideal gears is obtained under the form (Maatar & Velex, 1996), (Velex et al., 2011):

$$\frac{L(\tau)}{L_m} = 1 + 2\sum_{k=1}^{\infty} Sinc(k\varepsilon_\alpha) Sinc(k\varepsilon_\beta) \cos(\pi k(\varepsilon_\alpha + \varepsilon_\beta - 2\tau)) \quad (26)$$

with: $L_m = \varepsilon_\alpha \dfrac{b}{\cos\beta_b}$, average contact length

$Sinc(x) = \dfrac{\sin(\pi x)}{\pi x}$ is the classic sine cardinal function which is represented in Figure 9.

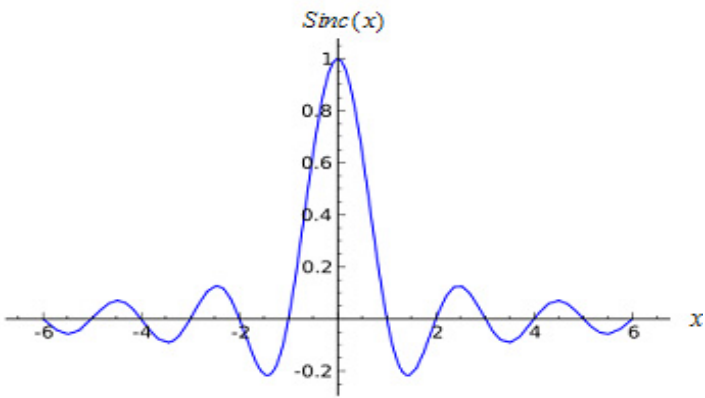

Fig. 9. Evolutions of $Sinc(x) = \dfrac{\sin(\pi x)}{\pi x}$.

The following conclusions can be drawn:

a. for spur gears, $\varepsilon_\beta = 0$ and $Sinc(k\varepsilon_\beta) = 1$
b. it can observed that the time-varying part of the contact length disappears when either $\varepsilon_\alpha$ or $\varepsilon_\beta$ is an integer
c. harmonic analysis is possible by setting $k = 1, 2, \ldots$ in (27) and it is possible to represent the contact length variations for all possible values of profile and overlap contact ratios on a unique diagram. Figure 10 represents the RMS of contact length variations for a realistic range of contact and overlap ratios. It shows that:
   - contact length variations are significant when $\varepsilon_\alpha$ is below 2 and $\varepsilon_\beta$ below 1
   - contact length is constant when $\varepsilon_\alpha = 2$ ($\varepsilon_\alpha = 1$ has to avoided for a continuous motion transfer) and /or $\varepsilon_\beta = 1$
   - for overlap ratios $\varepsilon_\beta$ above 1, contact length variations are very limited.

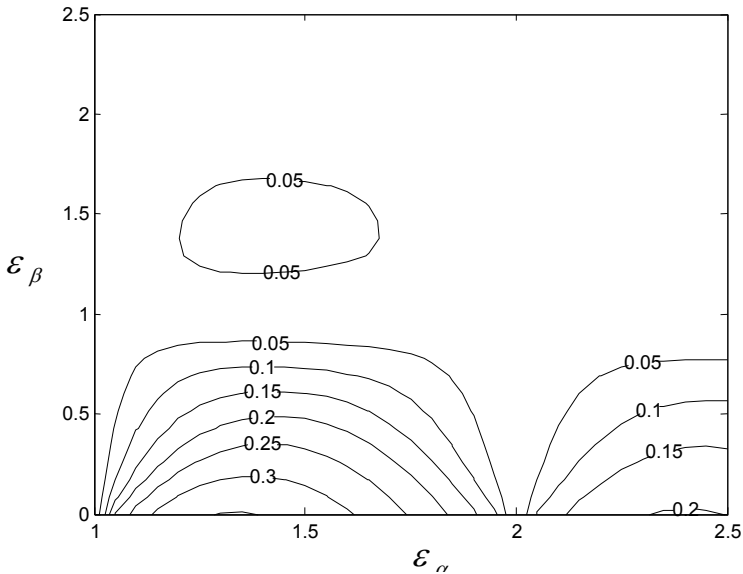

Fig. 10. Contour plot of the R.M.S. of $L(\tau)/L_m$ for a range of profile and transverse contact ratios.

### 4.3 Approximate expressions – Orders of magnitude

Mesh stiffness can be determined using the Finite Elements Method but it is interesting to have orders of magnitude or approximate values at the design stage. For solid gears made of steel, an order of magnitude of the mesh stiffness per unit of contact length $k_0$ is $\cong 1.3 \; 10^{10}$ N/m². More accurate expressions can be derived from the ISO 6336 standard which, for solid gears, gives:

$$k_0 \cong \cos\beta \frac{0.8}{q} \qquad (27)$$

with
$\beta$ : helix angle (on pitch cylinder)

$$q = C_1 + \frac{C_2}{Zn_1} + \frac{C_3}{Zn_2} + C_4 x_1 + C_5 \frac{x_1}{Zn_1} + C_6 x_2 + C_7 \frac{x_2}{Zn_2} + C_8 x_1^2 + C_9 x_2^2$$

coefficients $C_1,...,C_9$ have been tabulated and are listed in Table 1 below
$Zn_i = \frac{Z_i}{\cos^3 \beta}$ , $i = 1,2$ are the number of teeth of the equivalent virtual spur pinion ($i = 1$) and gear ($i = 2$).
$x_i$ , $i = 1,2$ , are the profile shift coefficients on pinion ($i = 1$) and gear ($i = 2$)

| $C_1$ | $C_2$ | $C_3$ | $C_4$ | $C_5$ | $C_6$ | $C_7$ | $C_8$ | $C_9$ |
|---|---|---|---|---|---|---|---|---|
| 0,04723 | 0,15551 | 0,25791 | -0,00635 | -0,11654 | -0,00193 | -0,24188 | 0,00529 | 0,00182 |

Table 1. Tabulated coefficients for mesh stiffness calculations according to ISO 6336.

## 5. Equations of motion – Dynamic behaviour

### 5.1 Differential system

The equations of motion for undamped systems are derived by assembling all the elemental matrices and forcing term vectors associated with the gears but also the supporting members (shafts, bearings, casing, etc.) leading to a parametrically excited non-linear differential system of the form:

$$[\mathbf{M}]\ddot{\mathbf{X}} + [\mathbf{K}(t,\mathbf{X})]\mathbf{X} = \mathbf{F}_0 + \mathbf{F}_1(t,\mathbf{X},\delta e(M)) + \mathbf{F}_2(t,\dot{\Omega}_{1,2}) \quad (28)$$

where $\mathbf{X}$ is the total DOF vector, $[\mathbf{M}]$ and $[\mathbf{K}(t,\mathbf{X})]$ are the global mass and stiffness matrices. Note that, because of the contact conditions between the teeth, the stiffness matrix can be non-linear (partial or total contact losses may occur depending on shape deviations and speed regimes). $\mathbf{F}_0$ comprises the constant nominal torques; $\mathbf{F}_1(t,\mathbf{X},\delta e(M))$ includes the contributions of shape deviations (errors, shape modifications, etc.); $\mathbf{F}_2(t,\dot{\Omega}_{1,2})$ represents the inertial effects due to unsteady rotational speeds

### 5.2 Linear behaviour – Modal analysis

Considering linear (or quasi-linear) behaviour, the differential system can be re-written as:

$$[\mathbf{M}]\ddot{\mathbf{X}} + [\mathbf{K}(t)]\mathbf{X} = \mathbf{F}_0(t) + \mathbf{F}_1(t,\delta e(M)) + \mathbf{F}_2(t,\dot{\Omega}_{1,2}) \quad (29)$$

The time variations in the stiffness matrix $[\mathbf{K}(t)]$ are caused by the meshing and, using the formulation based on structural vectors, the constant and time-varying components can be separated as:

$$[\mathbf{K}(t)] = [\mathbf{K}_0] + \int_{L(\tau)} k(M) \bar{\mathbf{V}}(\mathbf{M}) \bar{\mathbf{V}}(\mathbf{M})^T dM \quad (30)$$

where $\bar{\mathbf{V}}(\mathbf{M})$ is the extended structural vector: structural vector completed by zeros to the total number of DOF of the model

Using an averaged structural vector as in (18):

$$\bar{\mathbf{V}}_0 = \frac{1}{T_m} \int_0^{T_m} \bar{\mathbf{V}}(M) dt \qquad (31)$$

(30) can be simplified as:

$$\left[\mathbf{K(t)}\right] = \left[\mathbf{K}_0\right] + \int_{L(t)} k(M) dM \, \bar{\mathbf{V}}_0 \bar{\mathbf{V}}_0^T = \left[\mathbf{K}_0\right] + k_m(t) \bar{\mathbf{V}}_0 \bar{\mathbf{V}}_0^T \qquad (32)$$

The separation of the average and time-varying contributions in the mesh stiffness function as $k(t) = k_m(1 + \alpha \varphi(t))$ leads to the following state equations:

$$[\mathbf{M}]\ddot{\mathbf{X}} + \left[\left[\mathbf{K}_0\right] + k_m(1+\alpha\varphi(t))\bar{\mathbf{V}}_0\bar{\mathbf{V}}_0^T\right]\mathbf{X} = \mathbf{F_0} + \mathbf{F_1}(t, \delta e(M)) + \mathbf{F_2}(t, \dot{\Omega}_{1,2}) \qquad (33)$$

For most gears, $\alpha$ is usually a small parameter ($\alpha \ll 1$) and an asymptotic expansion of the solution can be sought as a straightforward expansion of the form:

$$\mathbf{X} = \mathbf{X}_0 + \alpha \mathbf{X}_1 + \alpha^2 \mathbf{X}_2 + \ldots \qquad (34)$$

which, when re-injected into (33) and after identifying like order terms leads to the following series of constant coefficient differential systems:

Main order:

$$[\mathbf{M}]\ddot{\mathbf{X}}_0 + \left[\left[\mathbf{K}_0\right] + k_m \bar{\mathbf{V}}_0 \bar{\mathbf{V}}_0^T\right]\mathbf{X}_0 = \mathbf{F_0} + \mathbf{F_1}(t, \delta e(M)) + \mathbf{F_2}(t, \dot{\Omega}_{1,2}) \qquad (35\text{-}1)$$

$\ell^{th}$ order:

$$[\mathbf{M}]\ddot{\mathbf{X}}_\ell + \left[\left[\mathbf{K}_0\right] + k_m \bar{\mathbf{V}}_0 \bar{\mathbf{V}}_0^T\right]\mathbf{X}_\ell = -k_m \varphi(t) \bar{\mathbf{V}}_0 \bar{\mathbf{V}}_0^T \mathbf{X}_{\ell-1} \qquad (35\text{-}2)$$

Interestingly, the left-hand sides of all the differential systems are identical and the analysis of the eigenvalues and corresponding eigenvectors of the homogeneous systems will provide useful information on the dynamic behaviour of the geared systems under consideration (critical speeds, modeshapes).

The following system is considered (the influence of damping on critical speeds being ignored):

$$[\mathbf{M}]\ddot{\mathbf{X}}_\ell + \left[\left[\mathbf{K}_0\right] + k_m \bar{\mathbf{V}}_0 \bar{\mathbf{V}}_0^T\right]\mathbf{X}_\ell = \mathbf{0} \qquad (36)$$

from which the eigenvalues and eigenvectors are determined. The technical problems associated with the solution of (36) are not examined here and the reader may refer to specialised textbooks. It is further assumed that a set of real eigenvalues $\omega_p$ and real orthogonal eigenvectors $\mathbf{\Phi_P}$ have been determined which, to a great extent, control the gear set dynamic behaviour.

Focusing on dynamic tooth loads, it is interesting to introduce the percentage of modal strain energy stored in the gear mesh which, for a given pair ($\omega_p$, $\mathbf{\Phi_p}$), is defined as:

$$\rho_p = k_m \frac{\mathbf{\Phi_p}^T \overline{\mathbf{V}}_0 \overline{\mathbf{V}}_0^T \mathbf{\Phi_p}}{\mathbf{\Phi_p}^T \left[ [\mathbf{K}_0] + k_m \overline{\mathbf{V}}_0 \overline{\mathbf{V}}_0^T \right] \mathbf{\Phi_p}} = v_{\Phi p}^2 \frac{k_m}{k_{\Phi p}} \qquad (37)$$

with $\quad v_{\Phi p} = \mathbf{\Phi_p}^T \overline{\mathbf{V}} = \overline{\mathbf{V}}^T \mathbf{\Phi_p}$

$k_m, k_{\Phi p}$: average mesh stiffness and modal stiffness associated with ($\omega_p$, $\mathbf{\Phi_p}$)

It as been shown (Velex & Berthe., 1989) that $\rho_p$ is a reliable indicator of the severity of one frequency with regard to the pinion-gear mesh and it can be used to identify the potentially critical speeds $\omega_p$ for tooth loading which are those with the largest percentages of modal strain energy in the tooth mesh. If the only excitations are those generated by the meshing (the mesh frequency is $Z_1 \Omega_1$), the tooth critical speeds can be expressed in terms of pinion speed as:

$$\Omega_1 = \omega_p / k Z_1 \qquad k = 1, 2, ... \qquad (38)$$

Based on the contact length variations and on the transmission error spectrum, the relative severity of the excitations can be anticipated.

*Remark:* The critical frequencies are supposed to be constant over the speed range (gyroscopic effects are neglected). Note that some variations can appear with the evolution of the torque versus speed (a change in the torque or load can modify the average mesh stiffness especially for modified teeth).

For the one DOF tosional model in Figure 4, there is a single critical frequency $\omega = \sqrt{k_m / \overline{m}}$ whose expression can be developed for solid gears of identical face width leading to:

$$\Omega_1 \cong \frac{\Lambda}{k} \frac{\cos \alpha_p}{M Z_1^2} \sqrt{\frac{b}{B}} \sqrt{\cos \beta_b} \sqrt{\varepsilon_\alpha} \sqrt{1 + u^2} \qquad (39)$$

where $k = 1, 2, ...$ represents the harmonic order; $\Lambda = \sqrt{\frac{8 k_0}{\pi \rho}}$ ($\rho$ is the density), for steel gears $\Lambda \cong 2\,10^3 \ ms^{-1}$; $M$ is the module (in meter); $B$ is the pinion or gear thickness (supposed identical); $b$ is the effective contact width (which can be shorter than $B$ because of chamfers for example); $u = \frac{Z_1}{Z_2}$, speed ratio.

### 5.3 Dynamic response

### 5.3.1 The problem of damping

Energy dissipation is present in all geared systems and the amount of damping largely controls the amplification at critical speeds. Unfortunately, the prediction of damping is still a challenge and, most of the time; it is adjusted in order to fit with experimental evidence. Two classical procedures are frequently employed:

a. the assumption of proportional damping (Rayleigh's damping) which, in this case, leads to:

$$[\mathbf{C}] = a[\mathbf{M}] + b\left[[\mathbf{K}_0] + k_m \overline{\mathbf{V}}_0 \overline{\mathbf{V}}_0^T\right] \tag{40}$$

with: $a, b$, two constants to be adjusted from experimental results

b. the use of (a limited number of) modal damping factors $\varsigma_p$:

The damping matrix is supposed to be orthogonal with respect to the mode-shapes of the undamped system with the averaged stiffness matrix such that:

$$\mathbf{\Phi}_\mathbf{p}^T [\mathbf{C}] \mathbf{\Phi}_\mathbf{p} = 2\varsigma_P \sqrt{k_{\Phi p} m_{\Phi p}} \tag{41-1}$$

$$\mathbf{\Phi}_\mathbf{p}^T [\mathbf{C}] \mathbf{\Phi}_\mathbf{q} = 0 \tag{41-2}$$

with:  $\varsigma_p$ : modal damping factor associated with mode $p$

$k_{\Phi p}, m_{\Phi p}$ : modal stiffness and mass associated with mode $p$

or introducing the modal damping matrix $[\mathbf{C}_\Phi]$:

$$[\mathbf{C}_\Phi] = diag\left(2\varsigma_P \sqrt{k_{\Phi p} m_{\Phi p}}\right), \quad p = 1, N \bmod \tag{41-3}$$

Following Graig (1981), the damping matrix can be deduced by a truncated summation on a limited number of modes $Nr$ leading to the formula:

$$[\mathbf{C}] = \sum_{p=1}^{Nr} \frac{2\varsigma_p \omega_p}{m_{\Phi p}} \left([\mathbf{M}]\mathbf{\Phi}_p\right)\left([\mathbf{M}]\mathbf{\Phi}_p\right)^T \tag{42}$$

with: $\omega_p = \sqrt{\dfrac{k_{\Phi p}}{m_{\Phi p}}}$

Regardless of the technique employed, it should be stressed that both (41) and (42) depend on estimated or measured modal damping factors $\varsigma_p$ for which the data in the literature is rather sparse. It seems that $0.02 \leq \varsigma_P \leq 0.1$ corresponds to the range of variation for modes with significant percentages of strain energy in the meshing teeth. The methods also rely on the assumption of orthogonal mode shapes which is realistic when the modal density (number of modes per frequency range) is moderate so that inter-modal couplings can be neglected.

### 5.3.2 Linear response

Based on the previous developments, the linear response of gears to mesh parametric excitations can be qualitatively assessed. Response peaks are to be expected at all tooth critical speeds and every sub-harmonic of these critical speeds because mesh stiffness time variations may exhibit several harmonics with significant amplitudes. Figure 11, taken from Cai and Hayashi (1994), is a clear example of such typical dynamic response curves when the gear dynamic behaviour is dominated by one major tooth frequency $\omega_n$ (and can be simulated by using the classic one DOF model). The amplifications associated with each peak depends on i) the excitation amplitude (Eq. (27) can provide some information on the amplitude associated with each mesh frequency harmonic) and ii) the level of damping for this frequency. For more complex gear sets, interactions between several frequencies can happen but, as far as the author is aware, the number of frequencies exhibiting a significant percentage of modal strain energy in the tooth mesh seems very limited (frequently less than 5) thus making it possible to anticipate the potential dangerous frequency coincidences for tooth durability.

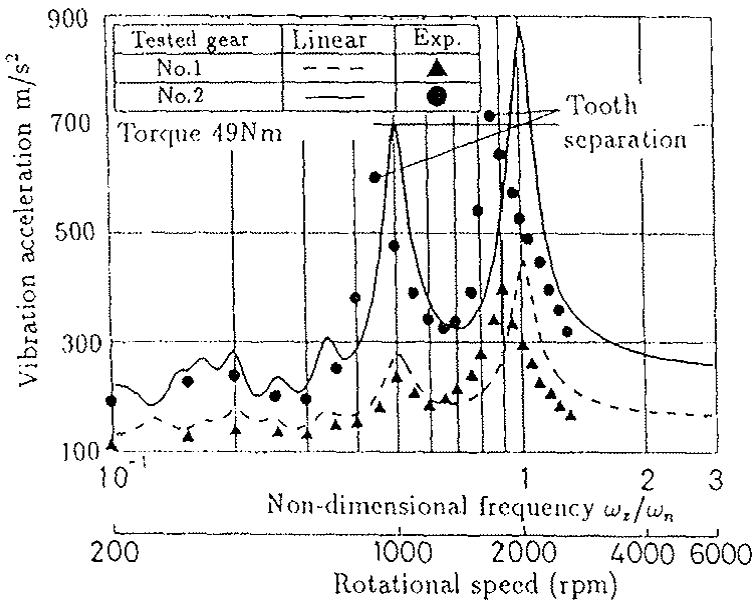

Fig. 11. Examples of dynamic response curves (Cay & Hayashi, 1994).

### 5.3.3 Contact condition – Contact losses and shocks

Only compressive contact forces can exist on tooth flanks and using (11), this imposes the following unilateral condition in case of contact at point *M*:

$$\mathbf{dF}_{1/2}(\mathbf{M}) \cdot \mathbf{n}_1 = k(M)\Delta(M)dM > 0 \qquad (43)$$

or, more simply, a positive mesh deflection $\Delta(M)$.

If $\Delta(M) \leq 0$, the contact at M is lost (permanently or temporarily) and the associated contact force is nil. These constraints can be incorporated in the contact force expression by introducing the unit Heaviside step function $H(x)$ such that $H(x) = 1$ if $x > 0$ and $H(x) = 0$ otherwise. Finally, one obtains:

$$\mathbf{dF}_{1/2}(\mathbf{M}) = k(M)\Delta(M)H(\Delta(M))dM\,\mathbf{n}_1 \qquad (44)$$

It can observed that contact losses are related to the sign of $\Delta(M) \cong \mathbf{V}_0^T\mathbf{X} - \delta e(M)$, from which, it can be deduced that two cases have to be considered:

a. $\delta e(M)$ is larger than the normal approach $\mathbf{V}_0^T\mathbf{X}$ which, typically, corresponds to large amplitudes of tooth modifications reducing the actual contact patterns, to spalls on the flanks (pitting) where contact can be lost, etc.
b. the amplitude of the dynamic displacement $\mathbf{X}$ is sufficiently large so that the teeth can separate ($\mathbf{X}$ is periodic and can become negative in some part of the cycle).

Momentary contact losses can therefore occur when vibration amplitudes are sufficiently large; they are followed by a sequence of free flight within the tooth clearance until the teeth collide either on the driving flanks or on the back of the teeth (back strike). Such shocks are particularly noisy (rattle noise) and should be avoided whenever possible. Analytical investigations are possible using harmonic balance methods and approximations of $H(x)$ (Singh et al., 1989), (Comparin & Singh, 1989), (Kahraman & Singh, 1990), (Kahraman & Singh, 1991), and numerical integrations can be performed by time-step schemes (Runge-Kutta, Newmark, etc.). The most important conclusions are:

a. contact losses move the tooth critical frequencies towards the lower speeds (softening effect) which means that predictions based on a purely linear approach might be irrelevant. The phenomenon can be observed in Fig. 11 where the experimental peaks are at lower speed than those predicted by the linear theory.

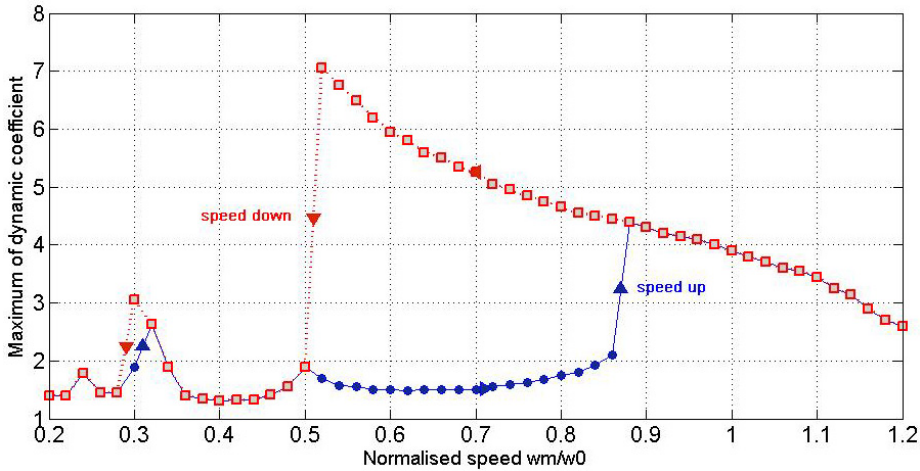

Fig. 12. Dynamic response curves by numerical simulations – Amplitude jumps – Influence of the initial conditions (speed up vs speed down), *2% of the critical damping, Spur gears* $Z_1 = 30$, $Z_2 = 45$, $M = 2mm$, *standard tooth proportions*.

b.  When contact losses occur, response curves exhibit amplitude jumps (sudden amplitude variations for a small speed variation),
c.  Because of a possibly strong sensitivity to initial conditions, several solutions may exist depending on the kinematic conditions i.e., speed is either increased or decreased
d.  damping reduces the importance of the frequency shift and the magnification at critical tooth frequency.

These phenomena are illustrated in the response curves in Figure 12.

## 6. Transmission errors

### 6.1 Definitions

The concept of transmission error (*TE*) was first introduced by Harris (1958) in relation to the study of gear dynamic tooth forces. He realised that, for high speed applications, the problem was one of continuous vibrations rather than a series of impacts as had been thought before. Harris showed that the measure of departure from perfect motion transfer between two gears (which is the definition of *TE*) was strongly correlated with excitations and dynamic responses. *TE* is classically defined as the deviation in the position of the driven gear (for any given position of the driving gear), relative to the position that the driven gear would occupy if both gears were geometrically perfect and rigid.

NB: *The concept embodies both rigid-body and elastic displacements which can sometimes be confusing.*

Figure 13 illustrates the concept of transmission error which (either at no-load or under load) can be expressed as angular deviations usually measured (calculated) on the driven member (gear) or as distances on the base plane.

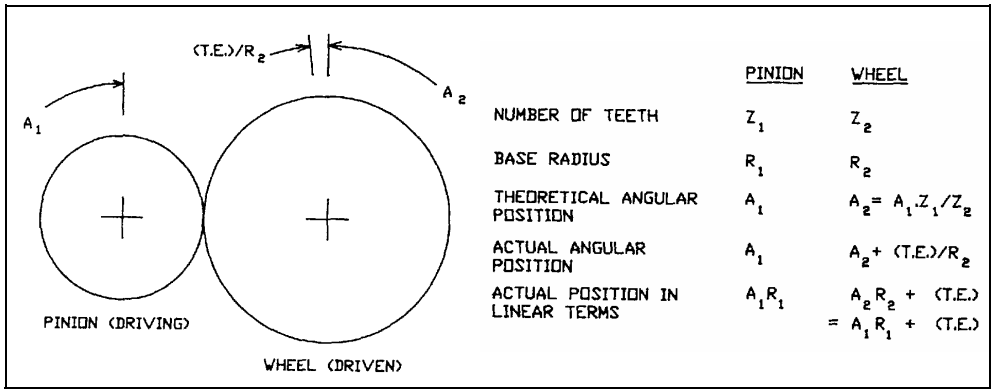

Fig. 13. Concept of transmission error and possible expressions (after Munro, (1989)).

Figures 14 and 15 show typical quasi-static *T.E.* traces for spur and helical gears respectively. The dominant features are a cyclic variation at tooth frequency (mesh frequency) and higher harmonics combined with a longer term error repeating over one revolution of one or both gears.

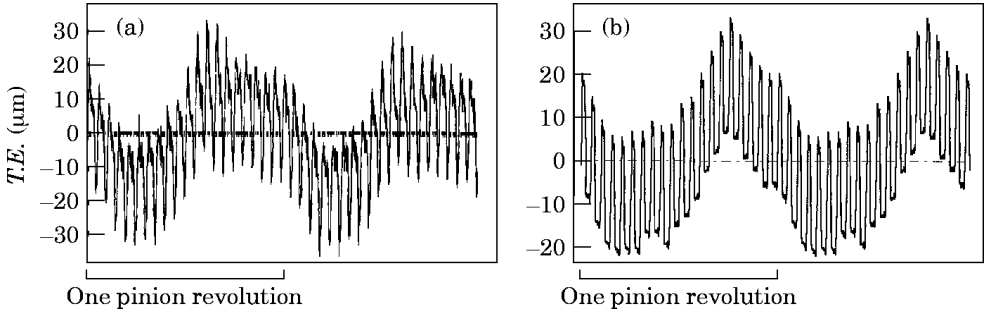

Fig. 14. Examples of quasi-static T.E. measurements and simulations – Spur gear (Velex and Maatar, 1996).

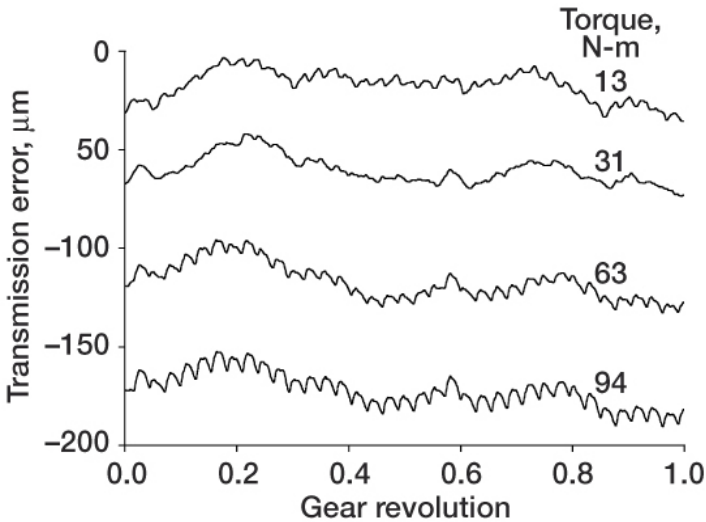

Fig. 15. T.E. measurements at various loads – Helical gear example.
NASA measurements from www.grc.nasa.gov/WWW/RT2001/5000/5950oswald1.html.

### 6.2 No-load transmission error (NLTE)

No-load *T.E.* (*NLTE*) has already been introduced in (2); it can be linked to the results of gear testing equipment (single flank gear tester) and is representative of geometrical deviations. From a mathematical point of view, *NLTE* is derived by integrating (2) and is expressed as:

$$NLTE = -\frac{E_{MAX}(t)}{\cos\beta_b} \tag{45}$$

### 6.3 Transmission errors under load

The concept of transmission error under load (TE) is clear when using the classic single degree of freedom torsional model (as Harris did) since it directly relies on the angles of torsion of the pinion and the gear. For other models (even purely torsional ones), the definition of *TE* is ambiguous or at least not intrinsic because it depends on the chosen cross-sections (or nodes) of reference for measuring or calculating deviations between actual and perfect rotation transfers from the pinion to the gear. Following Velex and Ajmi (2006), transmission error can be defined by extrapolating the usual experimental practice based on encoders or accelerometers, i.e., from the actual total angles of rotation, either measured or calculated at one section of reference on the pinion shaft (subscript *I*) and on the gear shaft (subscript *II*). *TE* as a displacement on the base plane reads therefore:

$$TE = Rb_1 \left[ \int_0^t \Omega_1 \, d\xi + \theta_I \right] + Rb_2 \left[ \int_0^t \Omega_2 \, d\xi + \theta_{II} \right] = Rb_1 \theta_I + Rb_2 \theta_{II} + NLTE \qquad (46)$$

with $\xi$, a dummy integration variable and $\theta_I, \theta_{II}$, the torsional perturbations with respect to rigid-body rotations (degrees of freedom) at node I on the pinion shaft and at node II on the gear shaft.

Introducing a projection vector **W** of components $Rb_1$ and $Rb_2$ at the positions corresponding to the torsional degrees of freedom at nodes I and II and with zeros elsewhere, transmission error under load can finally be expressed as:

$$TE = \mathbf{W}^T \mathbf{X} + NLTE \qquad (47\text{-}1)$$

which, for the one DOF model, reduces to:

$$TE = x + NLTE \qquad (47\text{-}2)$$

## 6.4 Equations of motion in terms of transmission errors

For the sake of clarity the developments are conducted on the one-DOF torsional model. Assuming that the dynamic contact conditions are the same as those at very low speed, one obtains from (21) the following equation for quasi-static conditions (i.e., when $\Omega_1$ shrinks to zero):

$$k(t,x) x_S = F_t + \zeta \cos \beta_b \int_{L(t,x)} k(M) \delta e(M) dM \qquad (48)$$

which, re-injected in the dynamic equation (21), gives:

$$\widehat{m} \ddot{x} + k(t,x) x = k(t,x) x_S - \kappa \frac{d^2}{dt^2}(NLTE) \qquad (49)$$

From (47-2), quasi-static transmission error under load can be introduced such that $x_S = TE_S - NLTE$ and the equation of motion is transformed into:

$$\widehat{m} \ddot{x} + k(t,x) x = k(t,x) [TE_S - NLTE] - \kappa \frac{d^2}{dt^2}(NLTE) \qquad (50)$$

An alternative form of interest can be derived by introducing the dynamic displacement $x_D$ defined by $x = x_S + x_D$ as:

$$\hat{m}\ddot{x}_D + k(t,x)x_D = -\hat{m}\frac{d^2}{dt^2}(TE_S) + (\hat{m}-\kappa)\frac{d^2}{dt^2}(NLTE) \tag{51}$$

The theory for 3D models is more complicated mainly because there is no one to one correspondence between transmission error and the degree of freedom vector. It can be demonstrated (Velex and Ajmi, 2006) that, under the same conditions as for the one DOF model, the corresponding differential system is:

$$[\mathbf{M}]\ddot{\mathbf{X}}_D + [\mathbf{K}(t,\mathbf{X})]\mathbf{X}_D \cong -[\mathbf{M}]\hat{\mathbf{D}}\frac{d^2}{dt^2}(TE_S) + \left[\frac{1}{Rb_2}\mathbf{I_P} + [\mathbf{M}]\hat{\mathbf{D}}\right]\frac{d^2}{dt^2}(NLTE) \tag{52}$$

where $\hat{\mathbf{D}} = k_m \cos\beta_b [\overline{\mathbf{K}}]^{-1} \overline{\mathbf{V}}$, $\mathbf{X}_D = \mathbf{X} - \mathbf{X}_S$, dynamic displacement vector

### 6.5 Practical consequences

From (51) and (52), it appears that the excitations in geared systems are mainly controlled by the fluctuations of the quasi-static transmission error and those of the no-load transmission error as long as the contact conditions on the teeth are close to the quasi-static conditions (this hypothesis is not verified in the presence of amplitude jumps and shocks). The typical frequency contents of *NLTE* mostly comprise low-frequency component associated with run-out, eccentricities whose contributions to the second-order time-derivative of NLTE can be neglected. It can therefore be postulated that the mesh excitations are dominated by $\frac{d^2}{dt^2}(TE_S)$. This point has a considerable practical importance as it shows that reducing the dynamic response amplitudes is, to a certain extent, equivalent to reducing the fluctuations of $TE_S$. Profile and lead modifications are one way to reach this objective. Equation (50) stresses the fact that, when total displacements have to be determined, the forcing terms are proportional to the product of the mesh stiffness and the difference between $TE_S$ and $NLTE$ (and not $TE_S$!). It has been demonstrated by Velex et al. (2011) that a unique dimensionless equation for quasi-static transmission error independent of the number of degrees of freedom can be derived under the form:

$$\cos\beta_b\, \hat{k}(t,\mathbf{X_S})\, TE_S^*(t) = 1 - \int_{L(t,\mathbf{X_S})} \hat{k}(M)e*(M)dM \tag{53}$$

with $\hat{A} = \frac{A}{k_m}$, $A^* = \frac{A}{\delta_m}$, for any generic variable *A* (normalization with respect to the average mesh stiffness and the average static deflection).

Assuming that the mesh stiffness per unit of contact length is approximately constant (see section 2-5), analytical expressions for symmetric profile modifications (identical on pinion and gear tooth tips as defined in Fig. 16) rendering $TE_S(t)$ constant (hence cancelling most of the excitations in the gear system) valid for spur and helical gears with $\varepsilon_\alpha \leq 2$ can be found under the form:

$$E = \frac{\Gamma \Lambda}{2\Gamma - 1 + \dfrac{1}{\varepsilon_\alpha}} \tag{54}$$

submitted to the condition $\Gamma \geq \dfrac{\varepsilon_\alpha - 1}{2\varepsilon_\alpha}$

with $E$: tip relief amplitude; $\Gamma$: dimensionless extent of modification (such that the length of modification on the base plane is $\Gamma \varepsilon_\alpha Pb_a$) and $\Lambda = \dfrac{Cm}{Rb_1 b k_0}$: deflection of reference.

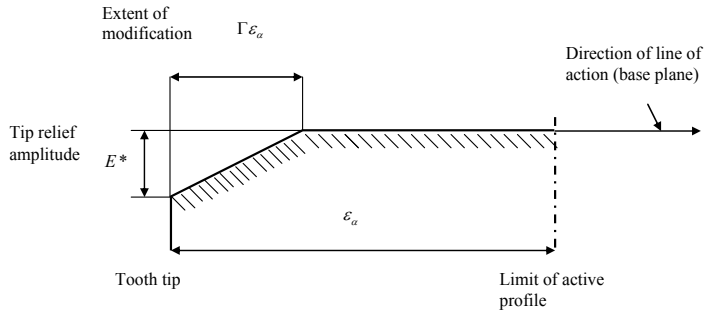

Fig. 16. Definition of profile relief parameters

Based on these theoretical results, it can be shown that quasi-static transmission error fluctuations for ideal gears with profile relief depend on a very limited number of parameters: i) the profile and lead contact ratios which account for gear geometry and ii) the normalised depth and extent of modification. These findings, even though approximate, suggest that rather general performance diagrams can be constructed which all exhibit a zone of minimum TE variations defined by (54) as illustrated in Figure 17 (Velex et al., 2011).It is to be noticed that similar results have been obtained by a number of authors using very different models (Velex & Maatar, 1996), (Sundaresan et al., 1991), (Komori et al., 2003), etc.

The dynamic factor defined as the maximum dynamic tooth load to the maximum static tooth load ratio is another important factor in terms of stress and reliability. Here again, an approximate expression can be derived from (51-52) by using the same asymptotic expansion as in (34) and keeping first-order terms only (Velex & Ajmi, 2007). Assuming that $TE_S$ and $NLTE$ are periodic functions of a period equal to one pinion revolution; all forcing terms can be decomposed into a Fourier series of the form:

$$-[\mathbf{M}]\hat{\mathbf{D}}\frac{d^2}{dt^2}(TE_S) + \left[\frac{1}{Rb_2}\mathbf{I_P} + [\mathbf{M}]\hat{\mathbf{D}}\right]\frac{d^2}{dt^2}(NLTE) = -\Omega_1^2 \sum_{n \geq 1} n^2 \left[A^*_n \sin n\Omega_1 t + B^*_n \cos n\Omega_1 t\right] \tag{55}$$

and an approximate expression of the dimensionless dynamic tooth load can be derived under the form:

$$r(t) = \frac{F_D(t)}{F_S} \cong 1 + \sum_p \sqrt{\rho_p \widehat{k}_{\Phi p}}\, Y_{pn}(t) \tag{56}$$

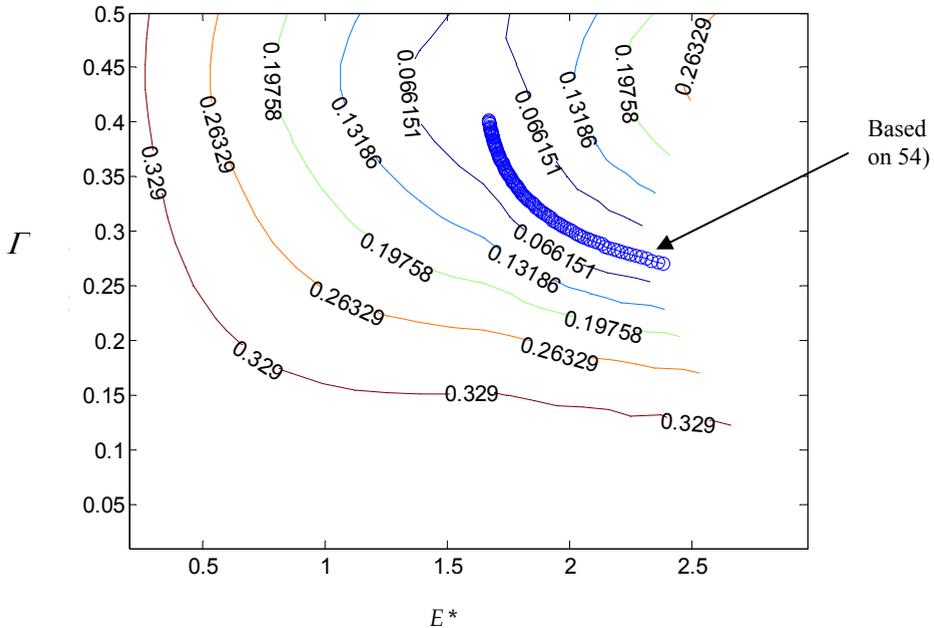

Fig. 17. Example of performance diagram: contour lines of the RMS of quasi-static transmission error under load - Spur gear $\varepsilon_\alpha = 1.67$.

with

$$Y_{pn}(t) = \sum_{n\geq 1} \frac{\bar{A}*_n\left[\left(\varpi_{pn}\right)^2 - 1\right] + 2\bar{B}*_n \varsigma_p\left(\varpi_{pn}\right)}{\left[\left(\varpi_{pn}\right)^2 - 1\right]^2 + 4\varsigma_p^2\left(\varpi_{pn}\right)^2} \sin n\Omega_1 t + \frac{\bar{B}*_n\left[\left(\varpi_{pn}\right)^2 - 1\right] - 2\bar{A}*_n \varsigma_p\left(\varpi_{pn}\right)}{\left[\left(\varpi_{pn}\right)^2 - 1\right]^2 + 4\varsigma_p^2\left(\varpi_{pn}\right)^2} \cos n\Omega_1 t$$

$$\varpi_{pn} = \frac{\omega_p}{n\Omega_1}$$

Equation (56) makes it possible to estimate dynamic tooth loads with minimum computational effort provided that the modal properties of the system with averaged stiffness matrix and the spectrum of $TE_S$ (predominantly) are known. One can notice that the individual contribution of a given mode is directly related to its percentage of strain energy in the meshing teeth and to the ratio of its modal stiffness to the average mesh stiffness. These properties can be used for identifying the usually limited number of critical mode shapes and frequencies with respect to tooth contact loads. They may also serve to test the structural modifications aimed at avoiding critical loading conditions over a range of speeds. It is worth noting that, since $\alpha$ is supposed to be a small parameter, the proposed methodology is more suited for helical gears.

# 7. Towards continuous models

## 7.1 Pinion, gear distortions

In the case of wide-faced gears, gear body deflections (especially those of the pinion) cannot be neglected and the torsion/bending distortions must be modelled since they can strongly affect the contact conditions between the teeth. For solid gears, one of the simplest approaches consists in modelling gear bodies by two node shaft finite elements in bending, torsion and traction as described in Ajmi and Velex (2005) which are connected to the same mesh interface model as that described in section 3 and Fig. 6. Assuming that any transverse section of the pinion or gear body originally plane remains plane after deformation (a fundamental hypothesis in Strength of Materials), gear bodies can then be sliced into elemental discs and infinitesimal gear elements using the same principles as those presented in section 2. The degrees of freedom of every infinitesimal gear element are expressed by using the shape functions of the two-node, six DOFs per node shaft element. By so doing, all the auxiliary DOFs attributed at every infinitesimal pinion and gear are condensed in terms of the degrees of freedom of the shaft nodes leading to a (global) gear element with 24 DOFs.

## 7.2 Thin-rimmed applications

The approach in 6.1 is valid for solid gears but is irrelevant for deformable structures such as thin-rimmed gears in aeronautical applications for example where the displacement field cannot be approximated by simple polynomial functions as is the case for shafts. Most of the attempts rely on the Finite Element Method applied to 2D cases (Parker et al., 2000), (Kahraman et al., 2003) but actual 3D dynamic calculations are still challenging and do not lend themselves to extensive parameter analyses often required at the design stage. An alternative to these time-consuming methods is to use hybrid FE/lumped models as described by Bettaieb et al, (2007). Figure 18 shows an example of such a model which combines i) shaft elements for the pinion shaft and pinion body, ii) lumped parameter elements for the bearings and finally iii) a FE model of the gear + shaft assembly which is sub-structured and connected to the pinion by a time-varying, non-linear Pasternak foundation model for the mesh stiffness. The computational time is reduced but the modelling issues at the interfaces between the various sub-models are not simple.

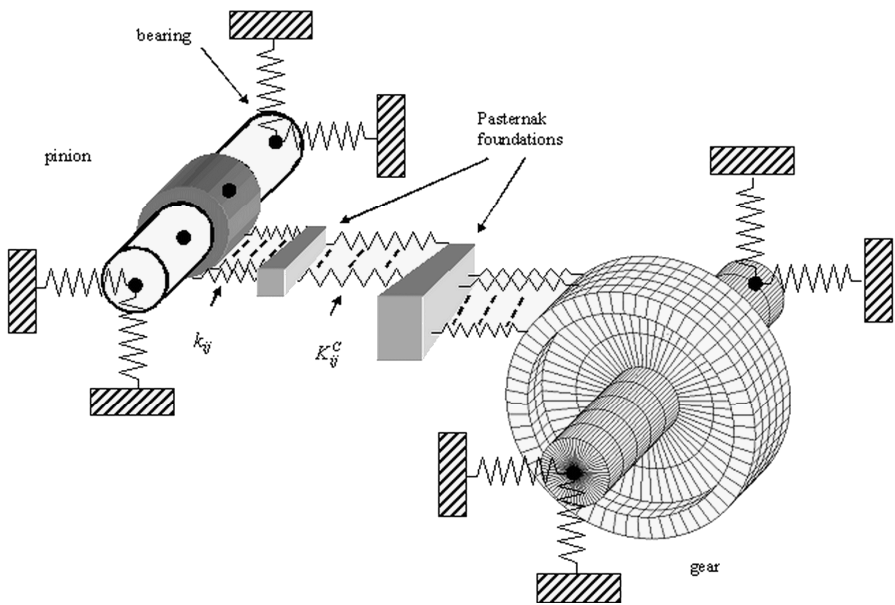

Fig. 18. Example of hybrid model used in gear dynamics (Bettaieb et al., 2007).

## 8. Conclusion

A systematic formulation has been presented which leads to the definition of gear elements with all 6 rigid-body degrees-of-freedom and time-varying, possibly non-linear, mesh stiffness functions. Based on some simplifications, a number of original analytical results have been derived which illustrate the basic phenomena encountered in gear dynamics. Such results provide approximate quantitative information on tooth critical frequencies and mesh excitations held to be useful at the design stage.

Gear vibration analysis may be said to have started in the late 50's and covers a broad range of research topics and applications which cannot all be dealt with in this chapter: multi-mesh gears, power losses and friction, bearing-shaft-gear interactions, etc. to name but a few. Gearing forms part of traditional mechanics and one obvious drawback of this long standing presence is a definite sense of déjà vu and the consequent temptation to construe that, from a research perspective, gear behaviour is perfectly understood and no longer worthy of study (Velex & Singh, 2010). At the same time, there is general agreement that although gears have been around for centuries, they will undoubtedly survive long into the 21st century in all kinds of machinery and vehicles.

Looking into the future of gear dynamics, the characterisation of damping in geared sets is a priority since this controls the dynamic load and stress amplitudes to a considerable extent. Interestingly, the urgent need for a better understanding and modelling of damping in gears was the final conclusion of the classic paper by Gregory et al. (1963-64). Almost half a century later, new findings in this area are very limited with the exception of the results of Li & Kahraman (2011) and this point certainly remains topical. A plethora of dynamic models can be found in the literature often relying on widely different hypotheses. In contrast, experimental results are rather sparse and there is certainly an urgent need for validated models beyond the classic results of Munro (1962), Gregory et al. (1963), Kubo (1978), Küçükay (1984 &87), Choy et al. (1989), Cai & Hayashi (1994), Kahraman & Blankenship (1997), Baud & Velex (2002), Kubur et al. (2004), etc. especially for complex multi-mesh systems. Finally, the study of gear dynamics and noise requires multi-scale, multi-disciplinary approaches embracing non-linear vibrations, tribology, fluid dynamics etc. The implications of this are clear; far greater flexibility will be needed, thus breaking down the traditional boundaries separating mechanical engineering, the science of materials and chemistry.